\documentclass[10pt,journal,letterpaper]{IEEEtran}
\usepackage{cite}
\usepackage[dvips]{graphicx}
\graphicspath{{./eps/}}
\DeclareGraphicsExtensions{.eps}
\usepackage{amsmath,amssymb,color}
\usepackage{algorithm}
\usepackage{algorithmic}
\usepackage{array}
\usepackage{mdwmath}
\usepackage{mdwtab}
\usepackage{eqparbox}
\usepackage{multirow}
\usepackage[tight,footnotesize]{subfigure}
\usepackage[hyphens]{url}
\usepackage{balance}
\usepackage{hyphenat}
\usepackage{wrapfig}
\begin{document}
%
%
\title
  {
  Technology in Hospitality Industry: Prospects and Challenges
  }
%
%
\author
  {
		\IEEEauthorblockN
    {
    Prasanna Kansakar,
    Arslan Munir, and
    Neda Shabani%
    }
  }

\maketitle
%
%
\begin{abstract}
The leisure and hospitality industry is one of the driving forces of the global economy. The widespread adoption of new technologies in this industry over recent years has fundamentally reshaped the way in which services are provided and received. In this paper, we explore some of the state-of-the-art technologies currently employed in the hospitality industry and how they are improving guest experiences and changing the hospitality service platform. We also envision some potential future hospitality services we can expect as the Internet of things (IoT) technology keeps growing. We recognize that the technological backbone of many hospitality establishments needs to be overhauled in order to facilitate the changing landscape of technology in the modern world. We discuss some fundamental challenges that need to be overcome to institute a lasting future-proof solution for the hospitality industry. We also touch upon the problems these challenges pose for guests and hospitality service providers (HSP). 
\end{abstract}
\sloppy
\nohyphens{
%
%
\section{introduction and motivation}
\label{section_introduction_and_motivation}

Domestic and international tourism has seen several years of steady growth. The revenue generated from accommodation, food and beverage, and other services provided to this large flux of travelers, has propelled the leisure and hospitality industry to become a key driver of the global economy. For sustained growth of this industry, experts in the field argue for major improvements in the type and quality of hospitality services to adapt to the changing consumption and travel behaviors of the evolving customer base. Specifically, these improvements are targeted towards attracting the new generation of technophile individuals traveling on a tight budget \cite{Deloitte_HospitalityOutlook_2014}. Implementation of these improvements compounds to a complete makeover of the service packages and the underlying technological framework currently used by hospitality service providers (HSP). The goal of these improvements should be: personalization of experiences and digitalization of services \cite{Deloitte_HospitalityOutlook_2014}.

Personalization of experiences is necessary to market services to individuals traveling on a limited budget. Personalization creates individualized guest experiences by incorporating flexibility and customizability to the offered service packages \cite{Deloitte_HospitalityOutlook_2014}. Most of the current packages marketed by HSP offer rigid and tailored experiences. These packages bundle different combinations of popular services in different price brackets with little to no means of negotiating adjustments. This leaves travelers to choose between all or nothing and they usually end up opting for the latter choice. If HSP have more flexible service package offerings, then guests can plan their experience according to their desires and their budgets. Crafting personalized value propositions for each guest requires a massive effort on both the guests' and the service providers' parts. This process can be simplified significantly by using an effective technological platform to manage the interaction between guests and service providers.

Digitalization of services is imperative to appeal to technophile guests. The goal of digitalization of services is to transition to a digital business model by pushing hospitality services to guests' touch-point \cite{Kasavana_HospitalityIndustry_2014}. 
A digital service platform affords guests the ability to browse, plan and pick activities at their own convenience thus facilitating seamless integration of technology into their travel experience. Booking and reservation services, location-based services and personalized communication, and social media integration are a few examples of digital services that entice technophile guests. There are a host of third party applications providing these services which guests are familiar with and rely upon. Revenue erosion to these third party applications and services is a growing concern to HSP \cite{Tossell_DigitalGuestStrategy_2015}. In order to compete with these third party applications, HSP must develop their own applications which provide better on-property and off-property services to guests. Through special incentives such as loyalty points, coupons and bonuses, guests can be encouraged to use in-house applications over third party applications. Providing digital services with the same quality as third party application services requires a sound technological infrastructure base with specialized computation and communication capabilities. This warrants the overhaul of current technological framework used by HSP.

The future of hospitality management industry is being shaped by the current boom in the Internet of things (IoT) technology. HSP must stay on the leading edge of IoT technology to maintain a competitive edge in the market. The IoT is the interconnection of everyday physical devices like sensors, actuators, identification tags, mobile devices, etc., such that they can communicate directly or indirectly with each other via local communication networks or over the Internet \cite{Munir_IFCIoT_2017}. The incorporation of IoT technology in the hospitality industry qualifies hotels as smart buildings which are important facets of smart cities \cite{Mohanty_SmartCities_2016}. The IoT paradigm offers HSP a nuanced means of interacting with guests and collecting their real-time data. This opens up new avenues for immediate, personalized and localized services as HSP can gauge guest behaviors and preferences with higher accuracy. The IoT also enables HSP to increase back-end efficiency of multiple departments \cite{Intelity_ForecastHotelTech_2016} (e.g. front desk, housekeeping, sales and marketing, etc.) as well as enact cost-saving policies like smart energy management \cite{Lee_Energy_2018} \cite{Hsiao_Energy_2018}. The IoT technology is already spreading through the hospitality industry with public terminals, in-room technologies and mobile applications \cite{Kasavana_HospitalityIndustry_2014} and some of the promising future IoT applications, such as body area sensor networks, environment monitoring and augmented reality experiences, will certainly usher in new business prospects. HSP should therefore aim to future proof their technology framework so that their systems can be easily upgraded in tandem with the changing IoT technological landscape.

Overall, the new technological upgrade of the hospitality industry should create a mutually beneficial platform by facilitating partnership between guests and HSP. The platform should ensure that guests are treated to an outstanding travel experience while also improving the operational and managerial efficiency for HSP. Furthermore, the new technological framework must be future proof; providing an easy upgrade schedule for addition of new/improved services. In this paper, we present a detailed overview of the role of technology in state-of-the-art hospitality services. We also describe potential future hospitality services following the burgeoning revolution of IoT technology. We then outline the challenges being currently being faced by HSP and discuss the need for overcoming these challenges to develop a lasting future-proof solution for the hospitality industry.

The remainder of this article is organized as follows. Section~\ref{section_state_of_the_art} describes the state-of-the-art hospitality services currently offered by providers. Section~\ref{section_future_services} envisions future services that guests can expect as hospitality industry continues to grow. Section~\ref{section_challenges} presents the major challenges and issues in designing solutions for the hospitality industry. Finally, Section~\ref{section_conclusions} concludes our article.
%
%
\section{state-of-the-art hospitality services}
\label{section_state_of_the_art}

HSP are making large IT expenditures to revamp their technological infrastructure base. In 2016, midscale hotels led in IT expenditure (7.3\%), trailed by upscale hotels (6.1\%) and luxury hotels (5.6\%)\cite{Intelity_ForecastHotelTech_2016}. The expenditures are largely focused on digitalization of the service platform to benefit both parties of the hospitality service exchange -- the guests and the service providers. Innovations in smart devices and IoT are driving the reform of technology used in the hospitality service platform. Guest interactions are being migrated towards on-screen and online interfaces through guest-facing systems which apart from being convenient for guests doubles as an opportunity for service providers to collect valuable data and feedback \cite{Deloitte_HospitalityOutlook_2014}. Digitalization, implemented by HSP through back-of-house (BoH) management systems, has helped improve operational efficiencies, enhance managerial effectiveness, reduce cost of goods sold, increase revenues and improve sustainability \cite{Kasavana_HospitalityIndustry_2014}.

In a digitalized hospitality service platform, guest-facing systems are the primary interfaces for interaction between the guests and the HSP. Therefore, it is imperative that these systems provide easy to use interfaces for guests to manage their travel experience. Guest-facing systems (shown in Figure \ref{figure_State-of-art}) include hospitality service mobile applications, point-of-sale (POS) terminals, hand-held devices, thin-client terminals, etc. \cite{Wang_Tech_2017} \cite{Ukpabi_Tech_2017}. These systems should be integrated seamlessly into all three phases of the guest cycle: pre-sale, point of sale, and post-sale phases so as to provide a complete digital service experience for the guests.

\begin{figure}[!t]
\centering
\includegraphics[width = 3.25in, bb = 14 13 635 799]{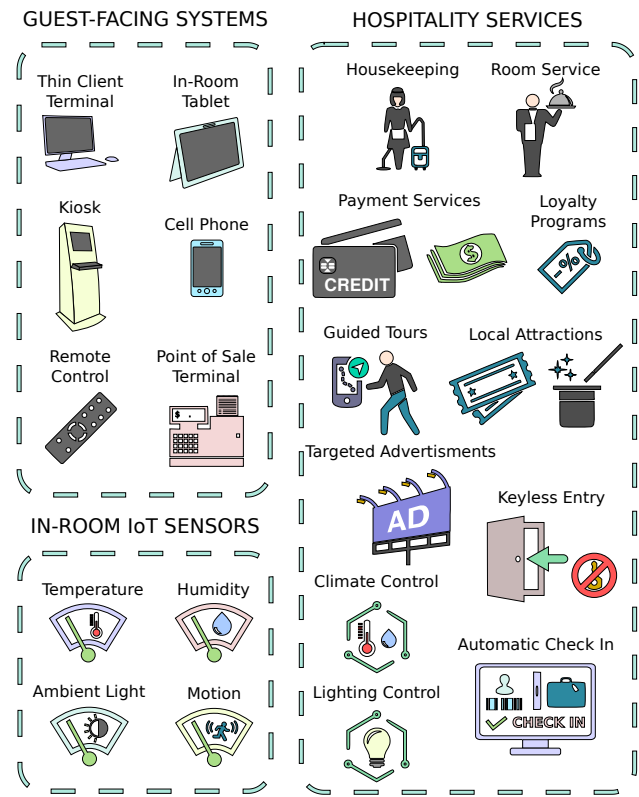}
\caption{State-of-the-art hospitality services}
\label{figure_State-of-art}
\end{figure}

Guest-facing systems improve guest experience in several different ways. Firstly, guest-facing systems ensure guest satisfaction by allowing guests to control their environment. Guest-facing systems empower guests with services such as automatic check-in and check-out services, keyless entry services, control of in-room functions etc. \cite{Wang_Tech_2017} (shown in Figure \ref{figure_State-of-art}). For example, Hilton and Starwood hotels offer guests automatic check-in and keyless entry service using their mobile apps \cite{DePinto_7TrendsIoTHospitality_2016}. Telkonet's EcoSmart Mobile offers similar mobile applications with the added features to allow guests to have control of in-room IoT products \cite{DePinto_7TrendsIoTHospitality_2016}. Samsung's Hotel Management Solutions and SINC entertainment solutions also allow guests to control in-room functions as well as check weather and flight information through a TV remote interface \cite{DePinto_7TrendsIoTHospitality_2016}. Hotels like Mondarian SoHo, The Plaza and The Marlin are placing tablets in their hotel rooms to provide guests with interfaces for controlling in-room functions \cite{VenturePact_HowIoTImprovesHospitality_2015}. Peninsula Hotels is developing their own line of proprietary in-room tablets which allow guests to order room service, message the concierge, arrange transportation, make free VOIP calls, and select TV stations and movies to stream onto the hotel room television \cite{Shallcross_MarriotEnseo_2016}.

\begin{figure*}[!t]
\centering
\includegraphics[width = 7in, bb = 13 15 1468 781]{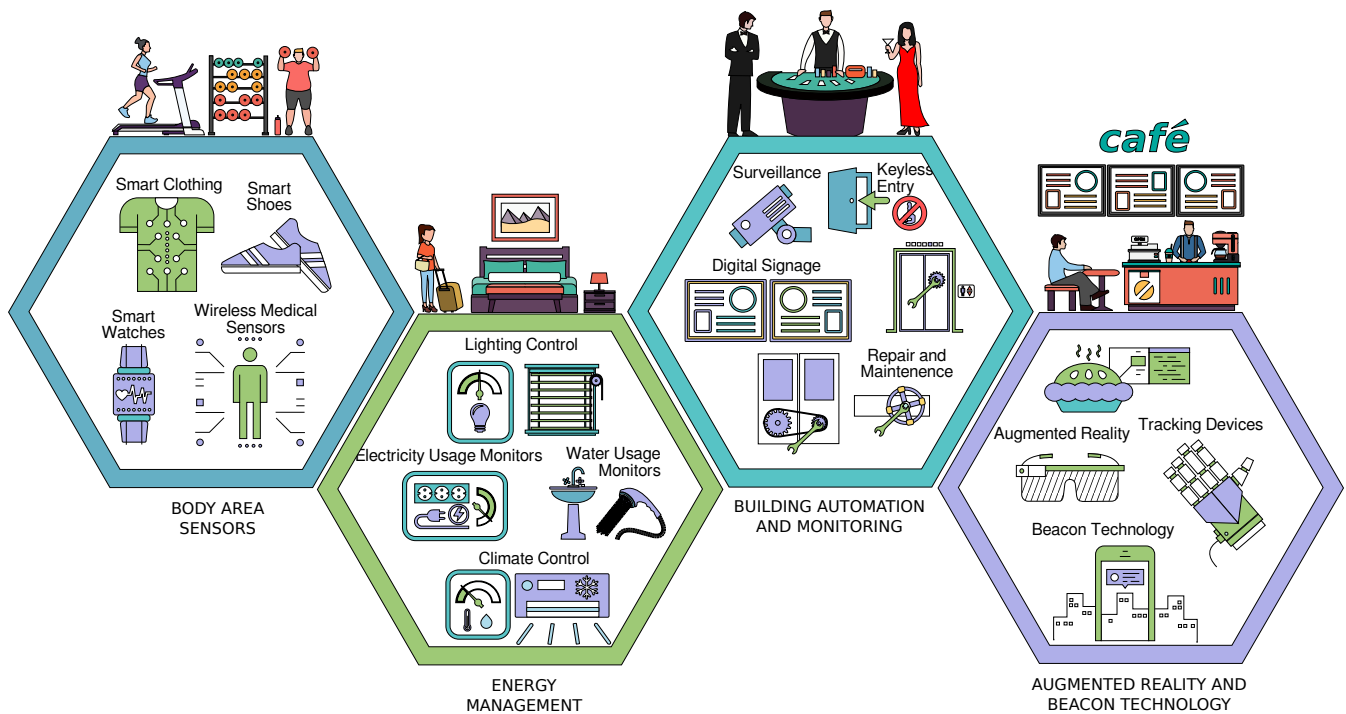}
\caption{Scope of future services in the hospitality industry}
\label{figure_IFCIoTFuture}
\end{figure*}

Secondly, guest-facing systems provide guests with location-based services which is another important service linked to guest satisfaction \cite{Tossell_DigitalGuestStrategy_2015}. More than 30 percent of hotels in 2016 allocated budgets for location-based technology \cite{DePinto_7TrendsIoTHospitality_2016}. Guest-facing systems enabled with location-based technology offer on-property and off-property guest services like digitally guided tours, recommendations of local events and attractions, as well as suggestions for dining and entertainment options (shown in Figure \ref{figure_State-of-art}). These services not only aid the guests in getting around and exploring during their stay, but, also enable service providers to keep guests within the revenue loop by preferably steering guests to sites and establishments that profit the HSP. For example, Fontainebleau Miami tailor their pre-arrival and checkout offers using their guests' location data \cite{DePinto_7TrendsIoTHospitality_2016}. Finally, guest-facing systems make it easy for guests to participate in loyalty programs with HSP \cite{Intelity_ForecastHotelTech_2016}. By using hotel loyalty mobile apps, guests can keep track of coupons and bonuses, and get notifications on deals and special offers.

The services offered to guests through guest-facing systems are driven by sophisticated BoH management systems. These systems are tasked with managing service staff and balancing operational costs and revenue without compromising quality of service provided to guests. The BoH management systems include property management system, customer relationship management, revenue and sales management, housekeeping maintenance software etc. \cite{Kasavana_HospitalityIndustry_2014}. The developments in guest-facing systems and IoT technology are significantly enhancing the capabilities of BoH management systems. For example, in-room IoT units like thermostats, motion sensors and ambient light sensors (shown in Figure \ref{figure_State-of-art}) can be used to control temperature and lighting in hotel rooms when they are unoccupied or unsold which can reduce energy costs by 20 to 45 percent \cite{DePinto_7TrendsIoTHospitality_2016}. Starwood Hotels and Resorts' ``daylight harvesting'' is such an energy-saving scheme which saves energy and increases indoor lighting consistency by automatically adjusting the energy-efficient LED lighting based on the natural light detected coming into the hotel room \cite{DePinto_7TrendsIoTHospitality_2016}.

The innovations in guest-facing systems are also reshaping the customer relations dynamic between guests and HSP. Guest-facing systems enable service providers to closely monitor the guest cycle by collecting data on specific guest preferences, behaviors and locations \cite{Wang_Tech_2017} \cite{Piccoli_Personalization_2017}. Service providers and BoH systems make use of this data to create custom guest profiles which they use to personalize service offers for repeat business. These custom guest profiles can be shared with a large network of partner service providers which ensures that services offered to guests are always highly personalized. Custom guest profiles also grant HSP the ability to entice guests into using their services by means of targeted advertisements and special insiders' guides and offers.

Another critical management task associated with BoH management systems is to build up the online brand value of HSP \cite{Lee_OnlineBrand_2014}. This includes developing and maintaining good customer relations through effective use of the social media platform, and engaging guests to rate and review services in online portals. The online standing of a company directly correlates to its revenue stream. Around 90\% of modern day technophile travelers base their decisions on online reviews when purchasing hospitality services. A single negative review can thus result in potential loss of a large number of customers. It is therefore necessary for BoH management systems to monitor online portals for bad reviews and ratings and take necessary action to mitigate their effects. However, through the effective use of social media platform such as live chat-based assistance for prompt response to guest queries, advertising group activities and services to a group of a guest's friends etc., hospitality services can be highly personalized which makes them lucrative in a guest's point of view.

The BoH management systems also help improve revenue per available room (RevPAR) \cite{Altin_RevPAR_2017} by speeding up the housekeeping and maintenance processes. By using in-room technologies and guest preference profiles, BoH management systems can schedule housekeeping services efficiently. This effectively reduces the downtime of hotel rooms, improves the utilization of labor resources, and significantly improves guest satisfaction. The use of housekeeping management systems and applications can help in reducing payroll costs by 10\% to 20\% \cite{Kasavana_HospitalityIndustry_2014}. The BoH management systems also help in maintenance of in-room and on-property smart systems. These systems help discover faults and failures in near real-time and thus facilitate prompt maintenance.
%
%
\section{scope of future hospitality services}
\label{section_future_services}

As the IoT ecosystem grows and spreads into different facets of everyday life, we can expect a future where every physical device that we use aggregates and analyzes our data and automatically provides us services. The hospitality industry is inclined to follow this growing trend to offer new types of services to its guests as well as to enact cost saving measures. In this section, we discuss some of potential services and use cases that the burgeoning IoT ecosystem may bring to the hospitality sector in the future. Figure \ref{figure_IFCIoTFuture} shows examples of IoT sensor and devices the different service categories they can be employed for.


\subsection*{Body Area Sensors}

Smart and wearable devices are at the forefront of the IoT revolution. Sales of devices such as smartphones, smart-watches, etc., are soaring and smart technology is beginning to be included in other wearable forms like smart clothing, smart shoes, etc. These devices gather user data like body temperature, heart rate, location, fitness activities etc. Wireless medical sensor technology further expands the scope of data collection by providing detailed data about organs and systems within the body. With proper analysis of data gathered through body area sensor networks, HSP can offer a host of new services to their guests such as, automatic adjustment of in-room temperature based on body temperature, adjustment of in-room lighting based on a guest's sleep-cycle, provide meal suggestions based on a guest's desired fitness goal, etc. HSP can also provide special facilities to guest's based on the type of medical devices they use. For example, service providers can filter out high carbohydrate and sugary meal options for diabetic guests, high cholesterol meal options for patients with heart disease, etc.


\subsection*{Augmented Reality and Beacon Technology}

HSP are coming up with new ways to incorporate augmented reality and beacon technology into their on-property systems. This technology can be used to provide guests with services such as digitally guided tours, previews of in-room environment (e.g., decor, facilities and amenities, etc.), immediate translation services for signs and other written materials, interactive restaurant menus with dish previews, critic reviews, food allergy information, etc. as well as interactive trivia games around on-property points of interest \cite{Tussyadiah_AugmentedReality_2017}. These services can be bundled as part in-house loyalty applications. As guests use these services, HSP can advertise new services or collect data to improve guest preference profiles \cite{HospitalityTech_BeaconsAugmentedReality_2015} \cite{Perey_AR_2015}.


\subsection*{Energy Management}

HSP can enact several cost-saving measures in the management of on-property energy consumption by leveraging IoT technology. These measures are particularly helpful in achieving "green" operation of on-property systems.
Some of the energy-saving systems currently in-place at many hotel properties include smart lighting and temperature control systems as well as use of low power devices like compact florescent bulbs, LED lights etc. IoT technology can significantly expand the scope of energy-saving systems. For example, IoT-enabled power outlets and IoT-enabled smart devices alert housekeeping and maintenance service personnel if a particular outlet exceeds a set limit for power consumption over a given period of time. The service personnel can then track down whether the guests are mindful of the power consumption or whether the power is leaking due to malfunctioning devices\cite{Lee_Energy_2018} \cite{Hsiao_Energy_2018}. IoT technology can also be employed to limit water consumption. This can be achieved through IoT-enabled smart bathrooms with smart shower heads, smart sinks, flow-controlled toilets etc.


\subsection*{Building Automation and Monitoring}

Both guests and service providers benefit from building automation and monitoring. New hospitality services such as keyless entry services, automated check-in and check-out services, digital concierge, etc. which will be brought about by developments in IoT-enabled systems will greatly improve guest satisfaction. These services are not only appealing for technophile users but, they can be specially helpful for guests with disabilities. Building automation also leads to greater operational and managerial efficiency for HSP. For example, in-room monitoring systems can be used to detect whether a room is occupied or unoccupied so as to schedule housekeeping services. IoT-enabled in-room and on-property guest-facing systems as well as other utility systems such as elevators, automated doors and windows, powerlines, pipelines, etc., can report faults and malfunctions and schedule preventive maintenance services before any problems are detected with regular physical inspections\cite{Vermesan_IoTForMaintenance_2014}.
%
%
\section{Challenges}
\label{section_challenges}

In this section, we identify four major challenges (Figure \ref{figure_Challenges}) associated with an effective IoT implementation in the hospitality industry. These challenges need to be addressed by the new technological infrastructures being adopted by HSP in order to sustain steady growth.

\begin{figure}[!t]
\centering
\includegraphics[width = 2.5in, bb = 14 13 461 527]{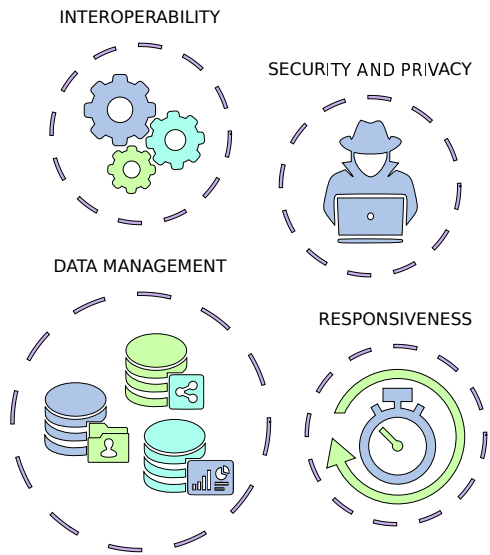}
\caption{Technological challenges in the hospitality industry}
\label{figure_Challenges}
\end{figure}


\subsection*{Interoperability}

The hospitality industry lacks standardization. Many HSP are developing their own proprietary solutions based on their own metrics and methodologies in order to accommodate the technological service demands of modern day guests\cite{SpecialNodes_HospitalityIoT_2013}. This has led to a diverse spectrum of implementations which are essentially targeted to provide a similar set of services. Although these implementations work well within the scope of a single property, they lack the potential to be extended to intra-organization and inter-organization scopes \cite{Wood_Hospitality_2013}. This imposes limitations on the usability of guest preference profiles on a broader scope because of the lack of a standard platform for sharing guest data across different businesses. This can lead to loss in potential revenue for HSP as they may be unable to effectively provide personalized services to their guests. Interoperability issues also impact guest experience as they create hassles and inconveniences that takes away from seamless user experience desired by guests. Non-standardized systems at different hotels introduce unwanted learning periods for guests during their stay. Such systems may also have issues in interfacing and using data from personal devices brought by guests. These problems warrant standardized vendor independent systems and solutions for hospitality industry.


\subsection*{Data Management}

Aggregation and analysis of guest data is an integral part of the hospitality service chain. With the introduction of new technologies and service platforms in the hospitality industry, data volume is bound to grow exponentially. Personalization of guest experience contributes significantly to increase in data volume. As personalized services become the norm in hospitality industry, HSP must treat each of their guests as unique individuals and maintain accurate and up-to-date records of their preferences and behaviors. HSP can collect guest data through guest-facing systems as well as personal guest devices connected to the hotel network. The BoH management systems in hotels must be capable of properly managing the influx of wide variety of guest data from wide variety of sources. In order to provide personalized services to guests, BoH management systems must analyze guest preference profile along with data about the state of the surrounding environment detected from IoT devices/sensors. This places a considerable computational burden on BoH management systems that can only be tackled through the use of specialized technological infrastructures. Additionally, secure sharing of relevant data from these guest profiles across different intra-organization and inter-organization systems is a monumental logistic challenge that requires both centralized and decentralized data management approaches.


\subsection*{Security and Privacy}

In order to provide guests with highly personalized services, it is necessary for HSP to track guest preferences, behavior, and location. HSP must ensure that guest data is used and stored properly so as to protect guests from physical, economic, and societal threats. Guest-facing systems and point-of-sale terminals are the most susceptible systems in hotels to security attacks. These systems should ensure that interactions with guests are secure and private by employing robust security measures to prevent data leaks and theft. Security primitives should also be supplemented in the hotel network for added security in interfaces with personal guest devices and in-room and on-property IoT devices. A secure hotel network prevents hackers from gaining access to guest data by attacking personal guest devices connected to the network. It also prevents hackers from reprogramming the hotel's IoT systems for annoying or malicious purposes. Adding strong security protocols in every guest interaction and every active connection on the hotel network requires significant computing resources. Moreover, these security protocols should be implemented close to the data source so that data is secured in as few number of hops in the network as possible. A decentralized computing platform is necessary to meet these requirements.


\subsection*{Responsiveness}

HSP must ensure prompt acknowledgement of guest requests and prompt delivery of services to guests. This can be achieved by digitalization of the interaction between guests and HSP. By pushing guest interactions to guest-facing systems and implementing automatic control through IoT sensors/devices, HSP can eliminate the need for human interaction and intervention when dealing with guests. These systems leave little room for miscommunication and confusion when interpreting guests' requests. These systems can also readily fulfill guests' requests faster than any dedicated hotel staff/personnel. This greatly improves responsiveness to guest requests and adds to the seamless experience desired by guests. Responsiveness is also crucial for a hotel's upkeep and maintenance. No or slow response to repair and maintenance needs can lower the hotel's revenue per available room (RevPAR). For example, a room cannot be rented if something as simple as the phone in the room is not working \cite{Altin_RevPAR_2017}. In hotels that have large scale IoT deployments, repair and maintenance requests can be responded to swiftly because most IoT sensors and devices can detect and self-diagnose problems. Timely repair and maintenance makes hotel rooms available for occupancy quickly thus reducing loss of revenue to maintenance. In order to improve responsiveness of hotel systems, they must be equipped with more computing resources and unfettered access to guest and BoH management data which requires a decentralized computing and data management platform.
%
%
\section{conclusions}
\label{section_conclusions}
In this paper, we outline many critical enhancements that need to be implemented in the hospitality industry to restructure its service platform to fit into the modern technological landscape. We identified personalization of experiences and digitalization of services as the two fronts in which these enhancements have to be focused. Many HSP have taken radical steps to remodel their services and we discuss some of these state-of-the-art hospitality services offered by them. We also envision several new future services that might be offered by the hospitality industry as some of the bleeding edge of systems, such as body area sensors, augmented reality, etc., enter maturity. We identify some fundamental challenges that need to be overcome to institute a lasting future-proof solution for the hospitality industry. We envision that future technological solutions for the hospitality industry will consist of geo-distributed systems that are capable of providing localized information and services, high volume data aggregation, security and privacy, and low latency event responses through energy efficient computing and bandwidth efficient communication resources. These solutions must also enable local, regional, and global analytics for providing valuable insights into improving quality of service as well as building better business models.


\vspace{8mm}

\noindent \textbf{Prasanna Kansakar} is a PhD student in the Department of Computer Science at Kansas State University, Manhattan, KS. His research interests include Internet of Things, embedded and cyber-physical systems, computer architecture, multicore, and computer security. Kansakar has an MS degree in computer science and engineering from the University of Nevada, Reno. Contact him at pkansakar@ksu.edu.

\vspace{8mm}

\noindent \textbf{Arslan Munir} is an assistant professor in the Department of Computer Science at Kansas State University, Manhattan, KS. His research interests include embedded and cyber-physical systems, computer architecture, multicore, parallel computing, fault tolerance, and computer security. Munir has a PhD in electrical and computer engineering from the University of Florida, Gainesville. He is a senior member of IEEE. Contact him at amunir@ksu.edu.

\vspace{8mm}

\noindent \textbf{Neda Shabani} Neda Shabani is a PhD student in the College of Human Ecology, Department of Hospitality Management at Kansas State University, Manhattan, KS. Her research interests include IT and technology in hospitality industry, such as, cybersecurity and privacy, augmented reality, Internet of things, computer architecture and big data. Shabani has BA in English Literature from Shiraz University, Iran and MS in Hospitality and Tourism Management from University of South Florida, Sarasota-Manatee. Contact her at nshabani@ksu.edu.


{
\balance
\bibliographystyle{IEEEtran}
\bibliography{IEEEabrv,IoT_Hospitality}
}

\fussy
} 

\end{document}